\documentclass[10pt,a4paper,oneside]{article}

\usepackage[utf8]{inputenc} 
\usepackage[T1]{fontenc}
\usepackage{lmodern}
\usepackage{url}            
\usepackage{booktabs}       
\usepackage{amsfonts}       
\usepackage{nicefrac}       
\usepackage[a4paper,left=3cm,right=2cm,top=2.5cm,bottom=2.5cm]{geometry}
\usepackage{graphicx}
\usepackage{tabularx}
\usepackage{amsmath}
\usepackage{braket}
\usepackage{amsthm}
\usepackage{amsbsy}
\usepackage{amssymb}
\usepackage{doi}
\title{Clamping effect on temperature-induced valence transition in epitaxial EuPd$_2$Si$_2$ thin films grown on MgO(001)}

\author{Sebastian Kölsch$^1$\,\footnote{Corresponding author. E-mail: \href{mailto:koelsch@physik.uni-frankfurt.de}{koelsch@physik.uni-frankfurt.de}} , Alfons Schuck$^1$, Olena Fedchenko$^2$, Dmitry Vasilyev$^2$, Sergeij\\Chernov$^2$, Lena Tkach$^2$, Christoph Schl\"uter$^3$, Thiago R. F. Peixoto$^3$, Andrii Gloskowski$^3$,\\ Hans-Joachim Elmers$^2$, Gerd Sch\"onhense$^2$, Cornelius Krellner$^4$, Michael Huth$^1$}
\date{%
    $^1$Thin films and nanostructures, Physical Institute, Goethe University Frankfurt,
Max-von-Laue Street 1, Frankfurt am Main 60438, Germany\\%
    $^2$Group Magnetism, Institute of Physics, Johannes Gutenberg University,\\
Staudingerweg 7, Mainz 55128, Germany\\
    $^3$Photon Science, Deutsches Elektronen-Synchrotron DESY, \\Notkestr. 85, 22607 Hamburg, Germany\\
    $^4$Crystal and Materials laboratory, Physical Institute, Goethe University Frankfurt,
Max-von-Laue Street 1, Frankfurt am Main 60438, Germany\\[2ex]%
    \today
}
\begin{document}
\maketitle
\vspace{-6mm}
\textbf{Abstract}
Bulk EuPd$_2$Si$_2$ show a temperature-driven valence transisition of europium from $\sim$+2 above 200\,K to $\sim$+3 below 100\,K, which is correlated with a shrinking by approximatly 2\% of the crystal lattice along the two a-axes.
Due to this interconnection between lattice and electronic degrees of freedom the influence of strain in epitaxial thin films is particularly interesting.
Ambient X-ray diffraction (XRD) confirms an epitaxial relationship of tetragonal EuPd$_2$Si$_2$ on MgO(001) with an out-of plane c-axis orientation for the thin film, whereby the a-axes of both lattices align. 
XRD at low temperatures reveals a strong coupling of the thin film lattice to the substrate, showing no abrupt compression over the temperature range from 300 to 10\,K.
Hard X-ray photoelectron spectroscopy at 300 and 20\,K reveals a temperature-independent valence of +2.0 for Eu.
The evolving biaxial tensile strain upon cooling is suggested to suppress the valence transition.
Instead low temperature transport measurements of the resistivity and the Hall effect in a magnetic field up to 5\,T point to a film thickness independent phase transition at 16-20\,K, indicating magnetic ordering.

\section{Introduction}
Besides being the most reactive lanthanide, the rare earth (RE) element Eu stands out in forming many intermetallic compounds in a divalent state, i.\,e. having a valence of nearly +2.
According to Hund's rule Eu$^{2+}$ (4f$^7$) has a strong magnetic moment connected to localized 4f-electrons, whereas Eu$^{3+}$ (4f$^6$) has zero moment \cite{sampathkumaran_new_1981}.
Despite the quite different electronic and magnetic configuration, both valence states can be rather close in energy.
As a consequence, in these compounds a variety of competing phenomena such as heavy-fermion like behaviour, antiferromagnetic ordering or a valence transition occur \cite{onuki_divalent_2017}. 
In particular, a transition of the valence state of Eu$^{2+}\rightarrow$ Eu$^{3+}$ may be tuned e.\,g. by temperature \cite{sampathkumaran_new_1981}, pressure \cite{adams_effect_1991} or high magnetic fields \cite{mitsuda_field-induced_1997}.
The valence transition is accompanied by a drastic change of the ionic radius of Eu by about 15\%.
Prototypical EuPd$_2$Si$_2$ is known to be a valence-fluctuating material with a temperature-induced valence change of Eu from 2.2 to 2.8 between 150-170\,K in a narrow temperature interval \cite{sampathkumaran_new_1981}.
This valence transition is accompanied by a large reduction ($\sim$2\%) of its a-lattice parameter upon cooling at ambient pressure \cite{onuki_divalent_2017}.
The c-lattice parameter instead remains mainly temperature-independent, whereby EuPd$_2$Si$_2$ already has a minimal Si-Si-bond length of 2.462\,\AA\, (ICSD 657595).
Recent research is focusing on the particularly strong coupling between electronic fluctuations and the lattice degrees of freedom near the second order critical endpoint in the p-T-phasediagram \cite{kliemt_strong_2022},
although it is not unambiguously clear whether EuPd$_2$Si$_2$ is on the high- or low-pressure side of this endpoint and how disorder or defects influence the valence transition.
Chemical substitution of, e.\,g., Pd with larger Au atoms, equivalent to applying negative chemical pressure, is claimed to lead to a first order phase transition in Eu(Pd$_{1-x}$Au$_x$)$_2$Si$_2$ with x $>$ 0.1 \cite{segre_valence_1982} such that the critical endpoint may become accessible via application of hydrostatic pressure.
As evident from resonant photoemission studies of EuPd$_2$Si$_2$ divalent Eu is found for the outermost Eu layers irrespective of temperature \cite{wertheim_final-state_1985}, suggesting different properties at the surface as compared to the bulk.
Ball milling EuPd$_2$Si$_2$ into the nanoparticle range leads to a broad distribution of valence transitions and possibly magnetic ordering at temperatures below 8\,K \cite{iyer_eu_2018}.
To our knowledge, up to now no thin film specific report regarding this ternary compound exists, leaving several questions regarding the coupling between lattice and electronic degrees of freedom unanswered.
Furthermore, for many RE-based compounds synthesis of single phase epitaxial thin films has not yet been achieved succesfully due to various difficulties during preparation \cite{chatterjee_heavy_2021}.
In this study we report experimental results concerning the growth of EuPd$_2$Si$_2$ epitaxial thin films and their temperature-dependent structural, magnetic and valence properties.

\section{Experimental Procedure}
Epitaxially grown EuPd$_2$Si$_2$ thin films were prepared on MgO(001) substrates using molecular beam epitaxy (MBE) in an ultra-high vacuum (UHV) chamber with a base pressure below $1\times10^{-10}$\,mbar.
During growth the pressure in the chamber did not exceed $1\times 10^{-8}$\,mbar.
The substrates are commercially available (Crystec GmbH) single-side epi-polished (R$_a <$  0.5 nm) pieces of MgO(001) with a size of $10\times10\times0.5\,$mm$^3$. 
Any chemical treatment was avoided to minimize exposure to water, which is known to hydroxylate MgO and thus reduce surface quality \cite{braun_situ_2020}. 
MgO crystallizes in the simple rocksalt structure (space group 225) with a cubic lattice parameter of a = 4.212\,\AA\, \cite{smith_low-temperature_1968} at room temperature. 
The MgO[001] surface reflects a square array of Mg- and O-ions, representing the most stable plane \cite{crozier_preparation_1992}. 
EuPd$_2$Si$_2$ has a tetragonal structure (space group 139) with lattice constants a = 4.231\,\AA\, and c = 9.86\,\AA\, (ICSD 657595). 
The substrate material was choosen because of the low misfit of $(a_{\text{EuPd}_2\text{Si}_2}-a_{\text{MgO}})/a_{\text{EuPd}_2\text{Si}_2} =  0.45\%$ at room temperature, implying a cuboid-on-cube growth with an out-of-plane c-axis orientation for the EuPd$_2$Si$_2$ thin film.\\

Before growth the substrates were degassed at 400$^\circ$C and thermally cleaned at 1000$^\circ$C under UHV for one hour, respectively, to desorb possible surface contaminations.
The temperature was then lowered to 450$^\circ$C for growth and allowed to stabilize. 
Sample temperature was measured indirectly via a radiatively coupled type-C thermocouple (95\%W/5\%Re - 74\%W/26\%Re) behind the heater.
For the codeposition of the low vapor pressure materials Si and Pd electron beam evaporators were used, whereas Eu was sublimated from a boron nitride crucible inside an effusioncell at a temperature of 470$^\circ$C measured at the bottom of the crucible. 
The evaporation rates of Si and Pd were continuously quantified throughout the deposition process and feedback-controlled via two water-cooled quartz microbalances (qmb) to ensure constant growth rates. 
For Eu deposition another qmb was used to check the rate directly before and after the deposition, showing no measurable deviation. 
During growth the temperature of the effusioncell was kept constant to better than 0.1$^\circ$C.
After growth the samples were allowed to cool to room temperature before an amorphous Si capping layer was deposited to prevent oxidation at atmosphere during further investigations at ambient pressure.
Between different growth steps the film surface properties were analyzed via reflection high-energy electron diffraction (RHEED) with a 15\,keV/5\,$\mu$A electron beam.
Scanning electron microscopy (SEM) images were aquired ex-situ to study the surface morphology in a FEI Nova NanoLab 600.
Furthermore the surface topography of the thin films were determined through atomic force microscopy (AFM), utilizing a Nanonis Nanosurf in non-contact mode under ambient conditions. 
For structural characterization high- and low-angle X-ray diffraction was done with a Bruker D8 Discover high-resolution diffractometer using Cu$_{K, \alpha}$ radiation with a parallelized primary beam and a diffracted-side monochromator in air.
To analyze small and high angle oscillations Bruker$^\prime$s DiffracPlus Leptos Software was used. 
Additional low temperature diffraction data was collected with a Siemens D500 diffractometer equipped with a Cryogenics closed-cycle helium refrigeration system down to 10\,K under high vacuum.
To study the transport properties, the thin films were patterned by means of UV-photolithography and low-energy Ar-ion etching to define 6-contact Hall bar structures with a cross-area of 30\,$\times$\,100\,$\mu \text{m}^2$. 
Low temperature magnetotransport measurements were then acquired via a variable temperature insert inside an Oxford helium flow cryostat between 3\,K and 300\,K in a magnetic field up to 5\,T.
For evaluation of the Eu mean valence Hard X-ray Photoelectron Spectroscopy (HAXPES) measurements of the Eu 3d core levels and the valence band at a photon energy of 3.4\,keV were conducted at beamline P22 of the storage ring PETRA III at DESY in Hamburg (Germany) using a time-of-flight momentum microscope \cite{babenkov_high-accuracy_2019}, \cite{medjanik_progress_2019}.
Typical Si-cap layer thickness was around 2\,nm for samples used for HAXPES and 6\,nm otherwise.
\section{Results and discussion}

\subsection{RHEED, AFM and SEM}
\begin{figure}[tb]
\begin{center}
\includegraphics[width=0.8\textwidth]{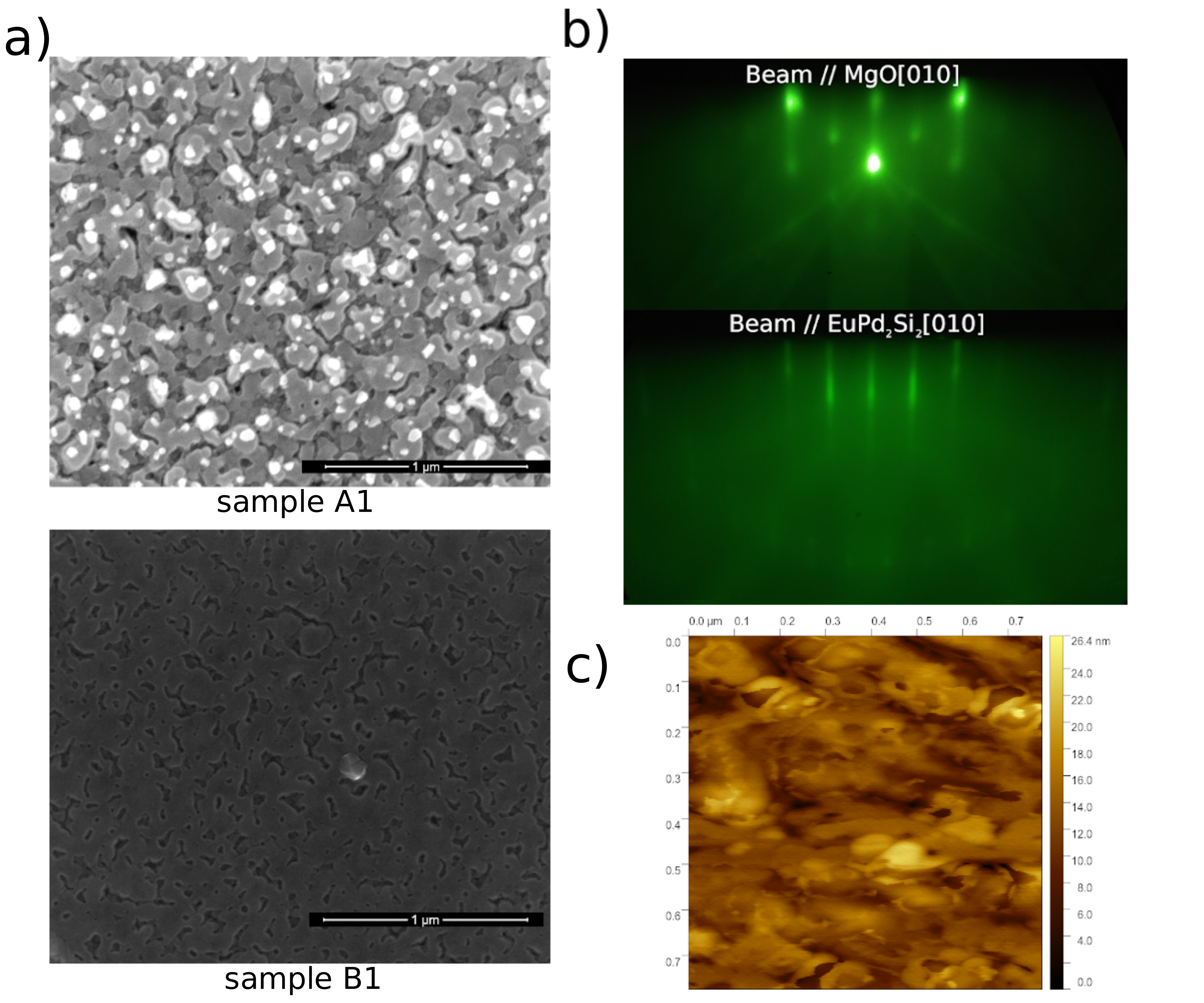}
\caption{\label{rheed} a) SEM image of a 24\,nm (A1) and 46\,nm thin film (B1) including Si capping layer. A small crystal with facettes is observed for the thicker film. The scale bar corresponds to 1\,$\mu$m. b) RHEED pattern before (top half) and directly after thin film deposition (bottom half) without any change of sample position. c) AFM image of sample B1 with added Si capping layer. The area of the micrograph is 770$\times$770\,nm$^2$. }
\end{center}
\end{figure}

In-situ RHEED azimuthal scans show a fourfold symmetry upon rotation around the surface normal of the pure MgO substrate after annealing at 1000$^\circ$C. 
After completion of the EuPd$_2$Si$_2$ thin film deposition again a fourfold symmetry of narrow streaks appear, see Fig.\,\ref{rheed} c. 
Comparison of the directions for the main symmetry axes from the substrate and the thin film implies a parallel alignment of their crystallographic a-axes. 

Even for thicker films broad Kikuchi bands arise, pointing to a laterally well ordered crystalline film.
In contrast the RHEED pattern vanishes completely after deposition of the Si capping layer indicating an amorphous overlayer structure with a thickness of some nanometers. 
SEM (Fig.\,\ref{rheed} a, b) and AFM (Fig.\,\ref{rheed} d) images of the Si-capped thin films reveal a locally smooth but stepped topography with a typical lateral island size in the order of 50-200\,nm and a mean surface roughness of $S_a \approx 3\,$nm.

\subsection{X-ray diffraction and reflectometry}
\begin{figure*}[tb]
\includegraphics[width=\textwidth]{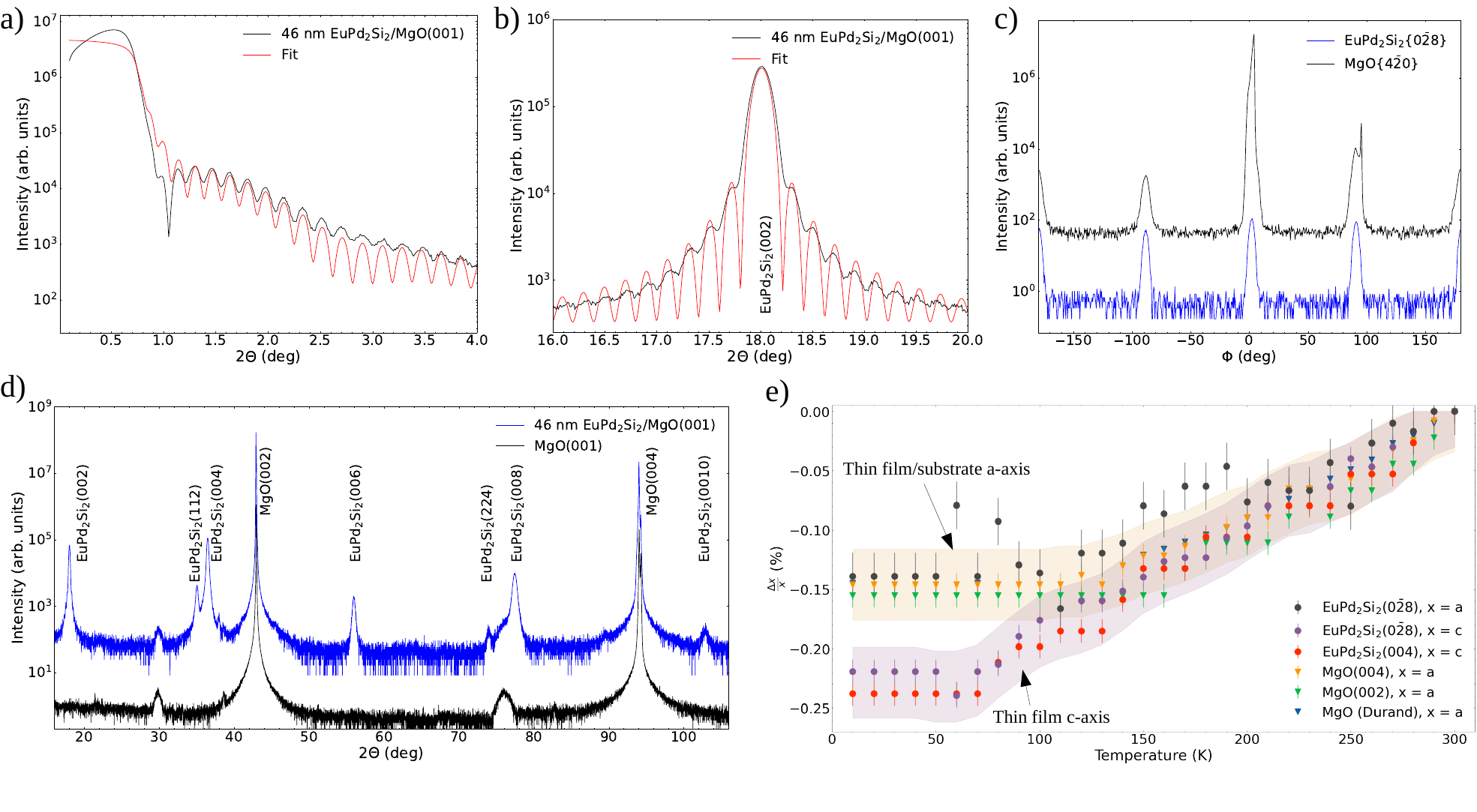}
\caption{\label{xrd} a) X-ray reflectometry scan (black) of a 46\,nm thin film (sample B1) with corresponding fit (red). b) Longitudinal symmetric X-ray scan near the EuPd$_2$Si$_2$(002)-reflex with fit of the Laue oscillations. c) $\phi$-scan around sample normal of MgO$\{4\bar{2}0\}$- and EuPd$_2$Si$_2\{0\bar{2}8\}$-reflexes in asymmetric geometry. d) Longitudinal symmetric scan of the same EuPd$_2$Si$_2$ thin film and a pure MgO(001) substrate. e) Temperature-dependent relative change of lattice parameters, $\Delta x/x$ for x=a and c, from cubic MgO and tetragonal EuPd$_2$Si$_2$ normalized at 300\,K (obtained from several reflexes as indicated in the legend). For comparison MgO single crystal data from Durand \cite{durand_coefficient_1936} are shown. Broad curves are a guide to the eye only.}
\end{figure*}
Despite of the measured roughness, Kiessig fringes arising from a relatively smooth interface and surface are visible in symmetric low angle X-ray reflectometry (XRR) scans (Fig.\,\ref{xrd} a) above 2$\Theta$ = 4$^\circ$, yielding an average film thickness $d$ of 46\,nm for EuPd$_2$Si$_2$ and approximately 2\,nm for Si, respectively, for sample B1 (see Table\,\ref{table} for a detailed comparison). 
Deviations between the experimental data and the fit may result from the oxidation of the Si capping layer leading to a partial SiO$_x$ overlayer, which is not included in the fitting procedure.
The symmetric high angle X-ray diffraction scan (Fig.\,\ref{xrd} d) shows (00$\ell$)-reflexes of EuPd$_2$Si$_2$ with even number $\ell$ appearing up to the 10th order, suggesting a well ordered epitaxial thin film. 
On the left hand side of the EuPd$_2$Si$_2$(008)-reflex a small shoulder is visible for all films, which is due to scattering from the sample holder and appears without any sample, too. 
At 35.0$^\circ$ an additional but small EuPd$_2$Si$_2$(112)-reflex appears, implying a minor contribution from misaligned crystallites as evident from the much higher structural form factor of the (112)-reflex as compared, e.\,g.\,, to the EuPd$_2$Si$_2$(004)-reflex.
For thicker films the higher order EuPd$_2$Si$_2$(224)-reflex at 74.0$^\circ$ starts to become visible. 
Laue oscillations next to the EuPd$_2$Si$_2$(002)- and (004)-reflex (Fig.\,\ref{xrd} b) indicate a high degree of structural order for the out-of-plane direction with a crystalline coherence length $L_c$ that is $\sim$90\% of the total layer thickness as obtained by XRR. 
Evaluating the precise position of the EuPd$_2$Si$_2$(0$\bar{2}$8)-reflex a complete relaxation of the in-plane lattice constant towards the bulk value of 4.237\,\AA\, of the single crystal is observed for all thicknesses investigated here.

$\phi$-scans in asymmetric reflection geometry of the MgO$\{4\bar{2}0\}$- and EuPd$_2$Si$_2\{0\bar{2}8\}$-reflexes show 4 regular spaced peaks at the same $\phi$-angles respectively, proving a parallel alignment of the corresponding a-axes.
The intensity of the MgO\{4$\bar{2}$0\}-reflexes in the asymmetric X-ray $\phi$-scan around the surface normal differs, which is caused by the high crystallinity and thus very sharp reciprocal lattice points in combination with a very small tilt offset of about 0.01$^\circ$ of the sample.
As a result, the normal vectors of the sample surface and the diffractometers $\phi$-circle are not exactly parallel, leading to an effective $\omega$-tilt during acquisition of the $\phi$-scan.
In the case of the EuPd$_2$Si$_2$\{0$\bar{2}$8\}-reflexes this effect is negligible, since these represent rather broad peaks, such that a change in the $\omega$-angle does only cause a minor change with respect to the measured intensity, as is directly evident from the much higher full width at half maximum (FWHM) visible in transverse scans (not shown).
The $\phi$-measurements thus confirm the epitaxial relationship deduced from RHEED to:

\begin{center}
MgO(100)$\parallel$EuPd$_2$Si$_2$(100) \& MgO[001]$\parallel$EuPd$_2$Si$_2$[001]
\end{center}

Extending the X-ray structural analysis to low temperatures yields further insight into the coupling between the two lattices and a clamping effect with respect to the in-plane lattice constant of EuPd$_2$Si$_2$. 
Using the asymmetric reflection geometry we track the position of selected reflexes in reciprocal space using iteratively coupled 2$\Theta/\omega$- and pure $\omega$-scans for each temperature to find the exact reflex position with maximum intensity.
Fig.\,\ref{xrd} e shows the resulting relative changes of the a- and c-axis lattice constants with temperature for the substrate and for a 46\,nm EuPd$_2$Si$_2$ thin film.
Due the low thermal expansion of MgO of less than $1\times 10^{-5}\,\text{K}^{-1}$ at room temperature \cite{smith_low-temperature_1968}, a maximum linear shrinkage of the substrate lattice by -0.14\,\% between 300\,K and 10\,K is expected. 
The measured temperature-dependent out-of-plane lattice constant of MgO\{001\} is thus in good agreement within the experimental error with values of the thermal expansion for MgO single crystals given in Ref.\,\cite{durand_coefficient_1936}. 
For the out-of-plane c-axis lattice constant of the EuPd$_2$Si$_2$ thin film we find a reduction of -0.23\% from both the (0$\bar{2}$8)- and the (004)-reflex, with a nearly linear decrease from 300\,K down to 70\,K.
Below 70\,K the c-axis lattice constant remains unchanged. 
The a-axis lattice parameter can be assessed by the EuPd$_2$Si$_2$(0$\bar{2}$8)-reflex, showing roughly the same thermal expansion as the out-of-plane cubic lattice constant of the MgO substrate. 
Note in particular that no abrupt change of the a-axis for EuPd$_2$Si$_2$ occurs, in contrast to the rapid reduction of $\sim$2\% observed below 150\,K for single crystals \cite{kliemt_strong_2022}.
As a consequence of the low thermal expansion of MgO and a strong coupling, i.\,e., a clamping effect of the thin film to the substrate, a high tensile biaxial in-plane strain is expected.

\subsection{HAXPES}
\begin{figure}
\begin{center}
\includegraphics[width=0.7\textwidth]{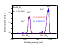}
\caption{\label{haxpes} Temperature dependence of the Eu 3d core spectra obtained with a photon energy $h\nu$ of 3.4 keV at 300\,K (red) and 20\,K (blue) for a 44\,nm thin film (sample B2). v represents the mean valence as derived from the ratio of the areas of the corresponding 3d core orbitals. The specified binding energies are referred to the Fermi level.}
\end{center}
\end{figure}
To investigate the electronic states in the EuPd$_2$Si$_2$ thin films in more detail, hard X-ray photoemission spectroscopy  experiments were conducted on a 44\,nm thin film with a 2\,nm Si capping layer (sample B2) with the aim to study the Eu mean valence v above and below the bulk phase transition temperature. 
Photoelectrons are excited with a photon energy of 3.4\,keV and the energy resolution is set to 200\,meV.
At this photon energy the inelastic mean free path of the photoelectrons is $\sim$4\,nm \cite{seah_quantitative_1979} and one obtains bulk-related information rather than surface properties.
Due to the spin-orbit interaction the Eu 3d spectrum is splitted into an Eu$_{5/2}$- and an Eu$_{3/2}$-component, whereby a chemical shift separates the Eu$^{2+}$- and Eu$^{3+}$-components by about 10\,eV in each case, allowing a precise determination of the mean valence from the corresponding peak areas \cite{mimura_temperature-induced_2011}.
The comparison between the high (300\,K) and low (20\,K) temperature spectra reveals no significant change with respect to the ratio of the Eu$^{2+}$-/Eu$^{3+}$-components and a temperature-independent valence near 2.0 (see Fig.\,\ref{haxpes}).
HAXPES measurements of the valence region (not shown) reveal the same density of states at 300\,K and at 20\,K and in particular an absence of Eu$^{3+}$ 4f states, confirming the presence of a pure Eu$^{2+}$ state even at 20\,K.
In contrast HAXPES measurements of polycrystalline bulk samples of EuPd$_2$Si$_2$ show a strong redistribution of the peak heights upon cooling below the valence transition temperature \cite{mimura_temperature_2004}.
Remarkably the temperature-induced valence transition is thus completly suppressed.

\subsection{Magnetotransport properties}
\begin{figure}
\includegraphics[width=\textwidth]{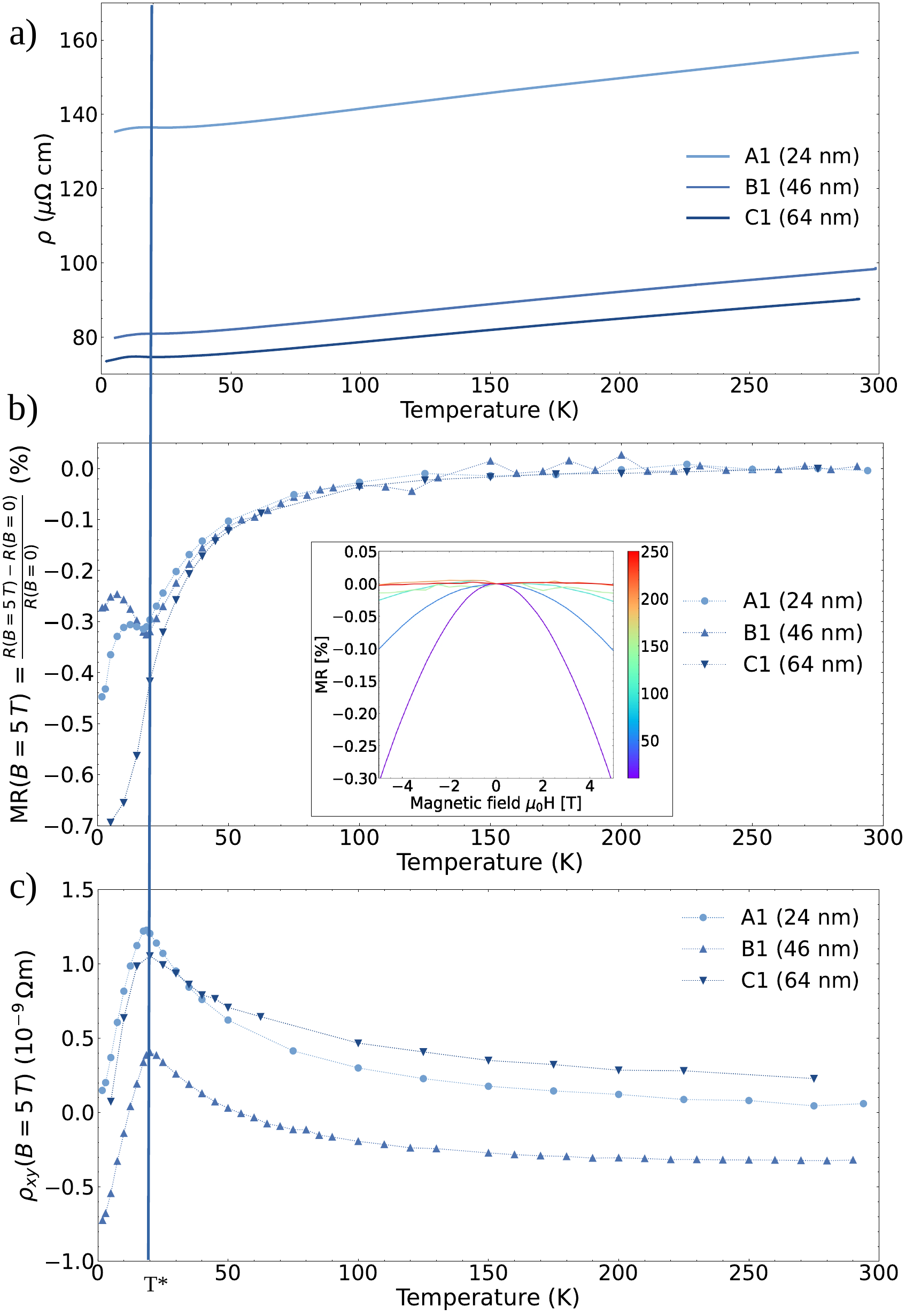}
\caption{\label{transport} a) Temperature dependence of the electrical 4-wire resistivity for three different films (samples A1, B1, C1) of thicknesses as indicated. b) Normalized transverse magnetoresistance measured at constant temperatures with typical magnetoresistance curves for sample B1 shown in the inset. c) Hall resistivity at 5\,T determined at constant temperature, see text for details.}
\end{figure}
All transport measurements (on samples A1, B1, C1, see Table\,\ref{table}) were done using a constant current of 100\,$\mu$A nearly parallel to the EuPd$_2$Si$_2$ a-axis, resulting in a current density of less than $2 \times 10^8\,\text{A/m}^2$.
The linear increase of the voltage with increasing current as observed at selected temperatures confirms the absence of heating effects.
The temperature dependence of the resistivity (Fig.\,\ref{transport} a) reveals a metallic behaviour of the EuPd$_2$Si$_2$ thin films above $\sim$50\,K.
At lower temperatures, a flat maximum around 16\,K with a small minimum around 20\,K appears, possibly indicating a phase transition.
This resistivity anomaly occurs at a nearly thickness-independent temperature $T^*$.
The residual resistivity ratio (RRR) between 280\,K and 3\,K amounts to $\sim$1.2 with a trend to larger values for thicker films.
With increasing film thickness the resistivity decreases strongly, which may be attributed to a more dense layer with stronger overlapping neighbouring islands, where the electrical transport is strongly influenced by grain boundary scattering.
In contrast, EuPd$_2$Si$_2$ single crystals containing small flux inclusions exhibit only a RRR slightly above 2 with a residual resistivity of approximatly 20\,$\mu\Omega$cm for current direction perpendicular to the c-axis \cite{kliemt_strong_2022}.
Application of a magnetic field perpendicular to the thin film leads to a very small negative magnetoresistance (MR) with a parabolic shape (inset of Fig.\,\ref{transport} b). 
The temperature dependence of MR(B = 5\,T) shows a minimum at the same characteristic temperature $T^*$ for samples A1 and B1.\\

Measurements of the Hall effect reveal a linear behavior with respect to the applied magnetic field at all temperatures and in magnetic fields up to 5\,T. 
For fixed magnetic field the temperature-dependent Hall resistivity exhibits a sample-independent maximum at $T^*$ (see Fig.\,\ref{transport} c). 
As to the reason for the observed anomaly we can only speculate at this stage, as we were so far not successful in acquiring temperature- and field-dependent magnetization data. 
Nevertheless, the sizable value for the Hall conductivity of, e.\,g., $\sim$20\,1/Ohm\,cm for sample C1, in conjunction with the suppressed valence transition and thus the constant charge carrier density, indicate a magnetic origin of the anomaly. 
As the magnitude of the magnetoresistance increases with decreasing temperature but does neither show saturation nor hysteresis effects as a function of magnetic field, a ferromagnetic ground state is unlikely.
A simple collinear antiferromagnetic ground state is, however, also unlikely, as an enhanced Hall effect due to topological reasons is not expected to occur \cite{smejkal_anomalous_2022}. 
However, even for a simple centrosymmetric crystal structure a topological Hall effect in the antiferromagnet state may appear, as has been experimentally shown recently for EuAl$_4$ \cite{shang_anomalous_2021}.

\begin{table}
\begin{center}
\caption{\label{table} Structural parameters and corresponding magnetotransport and valence properties of four samples. The deposition parameters for all samples are the same except for deposition time and therefor their total thickness $d$. na means ‘not available’.\\}
\begin{tabular}{cccccccc}
Sample & $d$/nm & $L_c$/nm & RRR & $\rho_{5\,K}$/$\mu\Omega cm$ & $T^*$ & v(20\,K)\\
\hline
A1 & 24 & 22 & 1.1 & 135.3 & 18 & na\\ 
B1 & 46 & 42 & 1.2 & 79.8 & 19 & na\\
B2 & 44 & 42 & na & na & na & 2.03\\
C1 & 64 & 61 & 1.2 & 74.0 & 19 & na\\
\end{tabular}
\end{center}
\end{table}

\section{Conclusion}
EuPd$_2$Si$_2$ grows as epitaxial and (001)-oriented thin film on Mg(001) substrates using molecular beam epitaxy with a typical lateral island size of the order of 100\,nm, while the film is fully coherent in the out-of-plane direction.
The evolution of a stepped surface indicates an island-like growth mode for EuPd$_2$Si$_2$ on MgO(001), which is typical for metal-on-insulator growth as, e.\,g., for Pd on MgO \cite{renaud_growth_1999}.
For all thin films investigated here a relaxed growth takes place, where the a- and c-axis lattice constants are equal to their bulk crystal values at room temperature.
A simple cuboid-on-cube model with the epitaxial relationship MgO(100)$\parallel$EuPd$_2$Si$_2$(100) and MgO[001]$\parallel$EuPd$_2$Si$_2$[001] was identified by RHEED and XRD experiments.
Due to a clamping-effect of the EuPd$_2$Si$_2$ thin film to the MgO substrate with negligible thermal expansion, the abrupt change of the a-lattice constant of EuPd$_2$Si$_2$ known from bulk material is suppressed, leading to a highly strained thin film upon cooling.
It is important to note, that the a-lattice parameter of EuPd$_2$Si$_2$ thin film shows the same temperature dependence as that of the substrate for all temperatures measured.
In contradistinction for the simple Eu/Nb/Al$_2$O$_3$-system a thickness dependence of the clamping temperature was found, where the Eu lattice is free to expand upon heating above a certain temperature.
Below this clamping temperature T$_{cl}$ the Eu in-plane lattice constants follow the behavior of the sapphire substrate, which may be explained by the temperature-dependent mobility of defects near the interface \cite{soriano_clamping_2004}.
For EuPd$_2$Si$_2$ thin films the valence transition is consequently suppressed, whereby a temperature-independent mean valence near 2.0 even at 20\,K is derived from HAXPES experiments.
Magnetotransport measurements indicate instead a phase transition between 16 and 20\,K reflecting a possible magnetic ordering.
The exact influence of the clamping towards, e.\,g., an antiferromagnetic ground state is still to be elucidated, but research along these lines is under way.

\section{Acknowledgment}
Special thanks are due to F. Ritter for assistance in collecting the low temperature x-ray data and T. Reimer (Gutenberg University Mainz) for argon ion beam etching of the lithographic patterned samples.
This work was supported by the German Research Foundation (DFG) via the TRR288 (422213477, projects A04, A03 and B04).
We thank the Bundesministerium f\"ur Bildung und Forschung (BMBF) through the project EffSpin-HAXPES (project number 05K16UMC) for additional funding.

\bibliographystyle{unsrt}

\end{document}